\documentclass[aps,prl,onecolumn,groupedaddress,showpacs,preprintnumbers,12pt]{revtex4}

\usepackage[dvips]{graphicx}
\usepackage{amsmath}
\usepackage{amsfonts,amssymb}

\usepackage{def}        

\begin{document}

\preprint{ULM-TP/02-7}
\preprint{August 2002}

\title{Comment on: ``Semiclassical theory of spin-orbit interactions using spin coherent states'' 
}

\author{Jens Bolte}
\email[]{jens.bolte@physik.uni-ulm.de}
\author{Rainer Glaser}
\email[]{rainer.glaser@physik.uni-ulm.de}
\author{Stefan Keppeler}
\email[]{stefan.keppeler@physik.uni-ulm.de}
\affiliation{Abteilung Theoretische Physik, 
        Universit\"at Ulm, 
        Albert-Einstein-Alle 11,
        D-89069 Ulm, Germany}


\begin{abstract}
We point out that a certain kind of combined classical translational and
spin dynamics -- claimed in 
[M.~Pletyukhov {\it et al.} Phys. Rev. Lett. {\bf 89} (2002) 116601]
to arise from the Pauli equation in the semiclassical limit $\hbar\to0$ 
for fixed spin -- only shows up if one simultaneously considers the 
high spin limit $S\to\infty$.
\end{abstract}


\maketitle

The main intention of the recent Letter \cite{PleAmaMehBra02} is to derive a
semiclassical trace formula for the density of states of quantum systems with
a spin-orbit interaction. The Hamiltonians under consideration are of the
form 
\begin{equation}
\label{Hamiltonian}
  \op{H} = H_0 (\hat{\vecq},\hat{\vecp}) 
        + \kappa \frac{\hbar}{2}\op{\vecs} \cdot 
        \vecC (\hat{\vecq},\hat{\vecp}) \, , 
\end{equation}
where $H_0 (\hat{\vecq},\hat{\vecp})$ describes the translational motion 
and $\op{\vecs}=(\op{s}_x,\op{s}_y,\op{s}_z)$ denotes normalized spin 
operators, i.e. generators of the $(2S+1)$-dimensional irreducible 
representation of the Lie algebra $\su(2)$. As their main tool to evaluate 
the trace of the time evolution operator $\ue^{-\frac{\ui}{\hbar}\op{H}t}$ 
the authors of \cite{PleAmaMehBra02} use coherent state path integrals for 
both translational and spin degrees of freedom. They claim that in the 
semiclassical limit $\hbar \to 0$, for fixed spin quantum number $S$, the 
classical periodic orbits entering the trace formula derive from coupled 
dynamics of the translational and the spin motion, including a back reaction 
of the spin on the translational part. This contradicts earlier results of 
two of us \cite{BolKep98,BolKep99a}. It is the aim of the present Comment 
to point out that the aforementioned kind of coupled classical dynamics can 
only appear in the combined limits $\hbar\to0$, $S\to\infty$ with 
$\hbar S = const$, but not for fixed spin $S$.

Two versions of semiclassical trace formulae for Hamiltonians of the 
type (\ref{Hamiltonian}) have previously been derived in \cite{BolKep99a}.
These two cases are distinguished by the way in which the semiclassical 
limit is performed. In the first scenario, referred to as the 
``weak-coupling limit'' in \cite{PleAmaMehBra02}, the single limit $\hbar\to0$
is considered, whereas the further parameters $\kappa$ and $S$ in 
(\ref{Hamiltonian}) are fixed. One then finds a trace formula which involves 
periodic orbits of the translational dynamics generated by the classical 
Hamiltonian $H_0(\vecq,\vecp)$ and solutions of the spin precession equation
$\dot{\vecs} = \kappa\,\vecC \times \vecs$ along these orbits. 
The second scenario of \cite{BolKep99a}, however, is concerned with the
combined limits $\hbar\to0$, $\kappa\to\infty$ such that 
$\tilde\kappa:=\kappa\hbar=const$ and therefore represents a 
``strong-coupling limit''. In this case the orbits relevant for the trace 
formula are determined by the ray Hamiltonians 
$H_0 + m \tilde\kappa |\vecC|$, $m=-S,\hdots,S$. The only remnants of the spin 
motion are certain geometric phases, but no classical spin precession enters. 
We also point out that only in the second scenario the problem of mode 
conversion can show up. These two scenarios are not properly distinguished 
in \cite{PleAmaMehBra02} as it is the first one which also applies to the 
Dirac equation, yet the second scenario proves the assumptions made in 
\cite{FriGuh93}. But most importantly, it should be noted that one can 
only achieve a {\it third} version of a trace formula, as intended in 
\cite{PleAmaMehBra02}, in a scenario different from the previous ones.
This would require to vary the additional parameter $S$ entering 
(\ref{Hamiltonian}) along with $\hbar$. 
Such an approach would provide a semiclassical theory with a different, 
rather than larger, range of validity.

The problematic step in the derivation of the trace formula in 
\cite{PleAmaMehBra02} is an incorrect application of the method of stationary
phase to the coherent state path integral. The latter is dominated by paths
that yield rapidly oscillating contributions to the total phase 
$\mathcal{R}/\hbar$ in the limit $\hbar\to 0$. Terms in the principal 
function $\mathcal{R}$ proportional to $\hbar$ hence must not be included 
in the determination of stationary points. 
If properly employed at this stage, the
two types of semiclassical limits studied in \cite{BolKep99a} will, as
it should be, reproduce the result obtained there.
In \cite{PleAmaMehBra02}
the authors, however, also include $\hbar$-dependent terms in the 
principal function $\mathcal{R}$ to
determine their classical paths. An inspection of the expression for 
$\mathcal{R}$ given in eqs. (7) and (8) in \cite{PleAmaMehBra02} reveals
that in order for this to result from a correct application of the method
of stationary phase the combined limits $\hbar\to 0$, $S\to\infty$ with
$\hbar S =const$ must be taken. This seems to have been overlooked in 
\cite{PleAmaMehBra02}, as also indicated by the fact that an example is 
given for the case $S=1/2$.

We remark that the trace formula searched for in \cite{PleAmaMehBra02} 
emerges as a special case, for the Lie group $G=\mathrm{SU(2)}$, of a
more general problem, 
see \cite{GuiUri90,TayUri92} and references therein. 
In these papers it is indeed proven that the combined limits 
$\hbar\to 0$, $S\to\infty$ with $\hbar S=const$ have to be considered 
for the desired trace formula to hold.

\bibliographystyle{my_unsrt}

\bibliography{literatur}

\end{document}